\documentclass[doublespacing]{elsart}

\usepackage{natbib}

\newcommand {\apgt} {\ {\raise-.5ex\hbox{$\buildrel>\over\sim$}}\ }
\newcommand {\aplt} {\ {\raise-.5ex\hbox{$\buildrel<\over\sim$}}\ }

\usepackage{graphicx}

\usepackage{amssymb}

\begin{document}

\begin{frontmatter}



\title{The Living Application: a Self-Organising System for Complex Grid Tasks.}


\author[aff:scs,aff:api,aff:lei]{Derek Groen}, 
\author[aff:api,aff:scs,aff:lei]{Stefan Harfst}, 
\author[aff:scs,aff:api,aff:lei]{Simon Portegies Zwart}

\address[aff:scs]{
  Section Computational Science, University of Amsterdam, Amsterdam, the Netherlands
} 
\address[aff:api]{
  Astronomical Institute "Anton Pannekoek", University of Amsterdam, Amsterdam, the Netherlands
}
\address[aff:lei]{
  Sterrewacht Leiden, Universiteit Leiden, Leiden, The Netherlands 
}

\pagebreak

Proposed running title: The living application

\author[aff:scs2,aff:api2]{Derek Groen}, 
\author[aff:api2,aff:scs2]{Stefan Harfst}, 
\author[aff:scs2,aff:api2,aff:lei2]{Simon Portegies Zwart}

\address[aff:scs2]{
  Section Computational Science, University of Amsterdam, Kruislaan 403, 1098 SJ, Amsterdam, the Netherlands
} 
\address[aff:api2]{
  Astronomical Institute "Anton Pannekoek", University of Amsterdam, Kruislaan 403, 1098 SJ Amsterdam, the Netherlands
}
\address[aff:lei2]{
  Sterrewacht Leiden, Universiteit Leiden, P.O. Box 9513, 2300 RA Leiden, The Netherlands 
}

Phone numbers and e-mail adresses:\\
Derek Groen: +31 20-5257582 (djgroen@uva.nl)\\
Stefan Harfst: +31 20-5257582 (harfst@science.uva.nl)\\
Simon Portegies Zwart: +31 20-5257510 (S.F.PortegiesZwart@uva.nl)

\pagebreak

\begin{abstract}
We present the living application, a method to autonomously manage
applications on the grid. During its execution on the grid, the living
application makes choices on the resources to use in order to complete
its tasks. These choices can be based on the internal state, or on 
autonomously acquired knowledge from external sensors.
By giving limited user capabilities to a living application, the
living application is able to port itself from one resource topology
to another. The application performs these actions at run-time without
depending on users or external workflow tools.
We demonstrate this new concept in a special case of a living
application: the living simulation. Today, many simulations require a
wide range of numerical solvers and run most efficiently if
specialized nodes are matched to the solvers. The idea of the living
simulation is that it decides itself which grid machines to use based
on the numerical solver currently in use.
In this paper we apply the living simulation to modelling the
collision between two galaxies in a test setup with two
specialized computers. This simulation switces at run-time between a
GPU-enabled computer in the Netherlands and a GRAPE-enabled machine
that resides in the United States, using an oct-tree N-body code
whenever it runs in the Netherlands and a direct N-body solver in the
United States.

\end{abstract}


\end{frontmatter}

\pagebreak

\section{Introduction} 
A grid application consists of a range of tasks, each of which may run
most efficiently using a different set of resources. Most of these
applications, however, use a fixed resource topology even though
certain tasks could benefit from using different resources. This can
be due to the computational demands of these tasks or due to a change
in resource availability over time. A wide range of work has been
done on developing external management systems that allow applications
to change grid resources during execution. This includes workflow
systems \citep{1239660,Kepler2006,Yu2005} or grid schedulers with migration capabilities
\citep{Condor2002,1080670} that support resource switches that are either part
of a predefined workflow or requested by the user.

An application management system that autonomously switches at run-time
has been proposed by \citet{1287574}, where a hierarchically
distributed application management system dynamically schedules and
migrates a bag-of-tasks style MPI application, using a static
hierarchy of schedulers to accomplish this. 

A self-adaptive grid application that does not require external
managers has been presented in \citet{1054349}. Although this
application does not use grid scheduling, it is able to autonomously
migrate to different locations and change its number of
processes. This has been accomplished by allowing all processes to
share knowledge and cooperate in managing the application's topology.

In this work, we propose the living grid application, in which the
application also decides where to run, and which is also able to
migrate itself at run-time to another computer when needed. The
intelligent migration from one computer to another can be realized
over a long baseline, but does not need to be designed this way (see
Sec.~\ref{sect:desc}). We then apply this method to a multi-scale
simulation and demonstrate its working on an intercontinental grid of
semi-dedicated computers by simulating the merging between two
galaxies, which provides a typical example for a multi-scale
simulation (Sec.~\ref{sect:exam}). In this simulation, we used
  a straightforward and autonomous resource selection scheme, where
  the optimal site is chosen from a predefined list of available
  resources. The simulation does not contain specific mechanisms to
  ensure fault tolerance or fault recovery.

\section{Living Application}
\label{sect:desc}
\subsection{Rationale}

A flexible approach is needed to execute a complex grid application
with multiple tasks and a diverse palette of resource requirements. 
The application should then be able to switch between tasks
at run-time and between the resources required for each of these
tasks, while maintaining the integrity of its data during these switches.

A switch requires the application to terminate its current execution,
output its current state, and from that reinitialize the application
using a new resource topology suited for the task at hand. Previously
this has been done on a grid only in orchestration with a workflow
manager.  A job submitted by a workflow manager lacks the ability to
change its resource topology during execution, as it does not have the
privileges to make use of grid schedulers. When running an application
with multiple tasks, this results in a 'bouncing' pattern where the
manager submits jobs which return once a switch is required, only to
be instantly submitted again to handle a different task. In the most
favorable case, the performance loss introduced by bouncing and
managerial overhead can be limited, but even then the successful
completion of the simulation depends on the availability of an
external manager, which is a potential single point of failure.

\subsection{How the living application works}

The living application switches between sites and tasks dynamically
and without external dependencies. It is based on four principles:

\begin{enumerate}

\item It makes decisions on which tasks to do and which resources to use.
\item It makes these decisions based on knowledge it has acquired at run-time.
\item It changes resources and switches between tasks.
\item It operates autonomously. 

\end{enumerate}

As a living application operates autonomously on the grid, it obtains
its privileges on its own without interacting with an external
workflow manager or user.

Upon initialization, the application is locally equipped with the
tools and data to perform the required tasks and the criteria for
switching between tasks or resource topologies. It is then submitted
as a job to the grid with the initial resource requirements defined by
the launcher. The living application begins execution on the grid and
continues to do so until either a switch or a termination is required.

The conditions for switching or termination are determined prior to the
start of the calculation or during run-time, but they are not
necessarily static. They can rely on the internal state of the
application, or on information from external sensors. When the
conditions for a switch have been met, the application will migrate to
different grid resources, switch to a different task, or both.

The switching between tasks requires two steps, which are finalizing
the old task (and any program it still uses) and starting up the new
task. During this switch, the application-specific data should be left
intact. The switching between sites requires a larger number of
actions, which are:

\begin{enumerate} 

\item Creating a set of files consisting of the current application,
  files with its parameters and data and a script that specifies the
  methods and conditions for switching and termination.

\item Creating a job definition for the application on the new resources.  

\item Authenticating (independently) on the grid.  

\item Transferring the files to the remote site (if this is not done 
  automatically by a resource broker).

\item Submitting the job, either through a resource broker or by 
  directly accessing the head nodes of grid sites.

\item Reinitializing the living application on the new site. 

\end{enumerate}

Additional file transfer may be required, if the application has
locally written data that is required elsewhere. The application
could initiate the
transfer of output files either during run-time (e.g. if
separate files are written) or just before a job terminates on one
machine (if data is appended to a single large file or data transfer
would cause overhead at run-time).

The living application requires some user privileges to initiate data
transfers and to autonomously migrate from one site to another. We obtain
these privileges by using a grid client interface to access a
credential management service. The details of this method are
discussed in Sec.~\ref{sect:sec}. The application requires access to the
grid client interfaces on all participating nodes to request these
privileges during execution. Once these privileges are granted, the
application can perform authentication, data transfers and job
submissions to the grid.

\subsubsection{Security Considerations}
\label{sect:sec}

User privileges on the grid are provided by an X.509 grid proxy
\citep{Welch04x.509proxy} which requires the presence of a
certificate, a private key and a correct pass phrase typed in by the
user. This proxy is represented by a temporary file with limited
lifetime. The easiest way to provide user privileges to a living
application would be to equip it with this file, transporting it as it
migrates, allowing it to reuse the proxy on remote locations. However,
this approach has three drawbacks:

First, the presence of a proxy file on a remote site poses a security
risk. If the file is not read-protected or stored in a shared account,
it may be possible for other grid users to copy the proxy. The
possession of this proxy enables them to impersonate the living
application user for the duration of the proxy's lifetime, providing
them with rights and resources that they could otherwise not use. Even
if the proxy is on a dedicated account and read-protected, local users 
with admin rights are able to copy it and use it for impersonation.

Second, it is not possible to cancel the application after the first
stage, as the proxy is initialized only at startup, after which it
travels around on remote sites. This may cause a malfunctioning
application to continue running and migrating until the proxy lifetime
is exceeded. An application that is equipped for self-reproduction may
iteratively spawns multiple successors which could lead to a grid
meltdown.

Third, for the same reasons as before it is also not possible to
prolong the lifetime of the proxy. This could cause the application to
terminate prematurely once the proxy lifetime is exceeded. Specifying an
excessively long lifetime relieves this problem, at the expense of
increasing exposure to the other two drawbacks.

To reduce these drawbacks we have chosen to use an intermediary
MyProxy server \citep{Basney2005} in our implementation. The user
initializes his or her proxy on the MyProxy server, which is encrypted
using a unique password. This password is stored in the living
application, which uses it to obtain short-lived user privileges from
the MyProxy server. If the password is stolen, others may be able to
get these short-lived privileges, but the user can remove access to
these privileges at any time by destroying the credential.

During application execution, the user can also extend the lifetime of
his MyProxy credential by renewing it. It is also possible to
replicate the credentials to other MyProxy servers, which allows the
application to use remote MyProxy servers if the local server has
died, rather than terminating itself upon switching.

\subsection{Living Simulation}

A special case of the living application is the living simulation.
Today, simulations of complex systems, in which the dynamic range
exceeds the standard precision of the computer, call for a wide range
of numerical solvers \citep{2008PCAA.book.....H}. Each of these
solvers may run most efficiently on a different computer
architecture. Most such simulations, however, are run on a single
computer even though they would benefit from running on a variety of
architectures. 

This can be solved by migrating the application at run-time from one computer
to another, in other words, by creating a living simulation. Such a
simulation loads the solvers as a library module and is able to probe the
internal variables of these solvers, making migration decisions based on this
information. We demonstrate the concept of the living application by applying
it to the (living) simulation of two galaxies merging.

The term living simulation has been previously defined as simulations
that fine-tune their behavior at run-time based on input from external
sensors, e.g. to provide input for performing
adaptive load balancing \citep{1360731}. In our definition we provide
the simulation with user privileges and expect it to function autonomously.
  
\section{Simulating galaxy mergers as a living simulation}
\label{sect:exam}

\subsection{Motivation}

A living simulation is based on the principle that it autonomously
switches between sites and solvers whenever required. This switching
is done dynamically and without external dependencies.  The simulation
is locally equipped with the required solvers, the switching criteria
and the initial conditions. It is then submitted as a job to the grid
with the initial resource requirements defined by the launcher. The
living simulation begins calculating on the grid and continues to do
so until either a switching condition or a termination condition has
been met.

By using the idea of the living applications, we have implemented and
tested a living simulation, in which the merger of two galaxies, each
with a central supermassive black hole (SMBH), is simulated. This is a
computationally expensive problem which requires integration with high
accuracy during close encounters and in the final stages of merging,
i.e. whenever the two SMBHs come close to each other. At an early
phase and at large separation of the two galaxies, however, less
accurate and therefore faster integration methods are sufficient. We
improve the performance and the dynamic range of the tree code
simulations (which are typically the method of choice for galaxy
merger simulations) by hybridizing the tree code with a direct
$N$-body solver.

\begin{figure}[t!]

  \centering
  \includegraphics[scale=0.3]{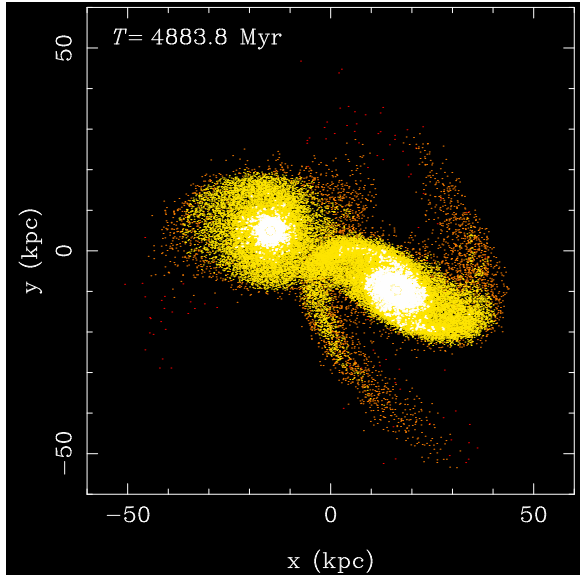}    
  \caption{Simulation snapshot of a 260k particle simulation,
    where the two galaxies approach for an initial interaction.}

\label{Fig:snapshot}
\end{figure}

In the scenario we are modelling, the two galaxies are initially well separated
by hundreds of kiloparsec, but they approach each other on a bound
orbit. Dynamical processes lead to a redistribution of energy and
momentum which causes, among other things, the formation of tidal
tails (see Fig~\ref{Fig:snapshot}). Eventually, these dynamical processes 
lead to the merger of the two galaxies.

In this merger, the two SMBHs, which reside in the galaxy cores, will
be brought close together until they form a binary SMBH. Modelling
the details of the formation of a binary SMBH and its subsequent
evolution requires a very accurate integration. Therefore, we choose
to switch from the tree code to a direct $N$-body solver at a
prespecified separation $r_a$ between the two SMBHs (see also
\cite{2008arXiv0807.1996P}). The switching allows us to follow the
full galaxy merger. This would not be possible using a single solver
due to the limited accuracy of the tree code and the computational
costs of the direct method.

In our living simulation, we make use of a dedicated GRAPE (GRAvity PipE,
\citet{1990Natur.345...33S}) special
purpose computer to perform direct-method
integration, and a graphics processing unit (GPU) to perform tree
simulations. The living simulation initially integrates using a tree
code on a GPU node, but switches to a direct integrator on a GRAPE
node when the separation between the two SMBHs $r_{\rm SMBH} <
r_a$. The simulation switches back to the GPU once $r_{\rm SMBH} \geq
r_a$.

\subsection{Implementation}
\label{sect:exam:impl}

We have used the Multiscale Software Environment (MUSE)\footnote{see
  http://muse.li} \citep{2008arXiv0807.1996P} package to conduct our
simulations. MUSE is a multi-scale/multi-physics astrophysical
framework that connects a variety of astrophysical codes, enabling
users to create combined simulations using Python scripts.  The
interfacing between existing solvers is realized using SWIG
\citep{1267513} with a uniformly defined interface for each domain. By
writing scheduling scripts, users are able to access the different
interfaces and create simulations that use multiple solvers for a wide
range of astrophysical problems.

The modular approach of MUSE lends itself very well to the grid
architecture. Modules run independently of each other and communicate
through the scheduling script. A grid-enabled scheduler would then
send each module to a different, suitable machine on the
grid. Furthermore, many astrophysical solvers run most efficiently on
dedicated and specialized computers. GRAPE boards, for example, have
been used extensively and very successfully in the field of stellar
dynamics
\citep[e.g.][]{2008ApJ...678..780G,2006ApJ...642L..21B,2003MNRAS.341..247B,2002ApJ...576..899P}.
In many cases, a MUSE application requires one or more specialized
platforms to run on and is therefore best run on a grid of such
specialized computers.

In previous work \citep{2008arXiv0807.1996P} we have extended MUSE
with a grid interface, allowing users to transfer files and perform
simulations on remote grid sites using a static and
  centralized scheduler which runs on the local user machine. The
grid interface has currently been implemented using the PyGlobus API
\citep{jackson02pyglobus}, and an alternative DRMAA-compliant
interface is under development.

Our test implementation consists of two components, a launcher to
initialize the living simulation and a job script that travels over
the grid during simulation. The launcher:

\begin{itemize} 
  \item Loads MUSE and the required modules, 
  \item reads the simulation input, 
  \item stores the parameters for each solver and the initial data for the first simulation stage, 
  \item transfers these files to the remote site, and
  \item submits the job script as a grid job to the remote site.  
\end{itemize}

The living simulation grid job executes the Python job script, which:

\begin{itemize}
  \item Initializes the simulation that will be used,
  \item reads and writes solver parameters and snapshots, 
  \item uses MUSE and SWIG to execute a simulation,
  \item transfers files, and
  \item submits a job script that computes the next simulation stage.
\end{itemize}

The job script is able to periodically check internal
  variables of the local solver at run-time using MUSE and
  SWIG. Consequently, the script is sensitive to changes in these
  variables, and autonomously performs actions (e.g. migration to a
  different site or file transfers) if certain conditions are met.

\subsection{Experiment setup}
For our experiments we make use of two grid nodes, one node equipped
with a GRAPE-6Af \citep{2005PASJ...57.1009F} at Drexel University in
Philadelphia, United States and one node with an Nvidia 8800
  Ultra GPU at the University of Amsterdam in the
Netherlands. The GRAPE-6Af has a peak performance of
  approximately 123 Gflops and an effective performance of up to $\sim$85
  Gflops when performing a direct-method simulation
  \citep{2005PASJ...57.1009F}.  The Nvidia 8800 Ultra has a
  theoretical peak performance of about 384 Gflops and a sustained
  performance of up to $\sim$100 Gflops when performing a N-body
  tree simulation using octgrav (E. Gaburov, personal communication).
The specification of the nodes can be found in Table~\ref{Tab:G3}. On
both nodes we have installed Globus 4.0.6 grid middleware
\citep{Foster2006} with GRAM, GridFTP and a MyProxy client, as well as
the MUSE framework. The nodes are linked using a regular internet
connection for which we have measured a latency of 100ms and a
bandwidth of approximately 550 kB/s.

On these nodes we run galaxy collision simulations (using
  simplified galaxy models, see below) that each last for 20 N-body
time units \citep{1986LNP...267..233H}. In all our runs, this duration
was sufficient to perform a full collision between the two galaxies.

The initial conditions for the galaxy collision consist of two
equally-sized Plummer sphere particle distributions
\citep{1911MNRAS..71..460P}, each of which has a central SMBH. We
perform simulations with $N=2k$ to $64k$ particles\footnote{i.e. 1024
  to 32768 particles per galaxy as well as 2 SMBH particles.}. The
total mass of particles in each galaxy is $M = 1$ and the mass of
individual particles is $m = M/N$. The SMBHs have each a mass of
$m_\mathrm{BH}=0.01$ or 1\% of the stellar mass of the galaxy.

When the two galaxies are far apart we use the tree-code
\citep{1986Natur.324..446B} in which further away particles are
grouped together to enable a hierarchical reduction in the force
computation.  The equations of motion are solved using the $2^{nd}$
order leap-frog particle integration scheme
\citep{1988csup.book.....H} with a fixed time step. The octgrav tree code we
use is written to run on a graphical processing unit
(Gaburov et. al., 2009, in preparation). The opening angle for the tree code is $\theta =
0.7$ and we use a time step of 1/64 N-body time unit (1/128 for the
largest data set). The direct-method integration is performed using
phiGRAPE \citep{2007NewA...12..357H}. In phiGRAPE, particles have
individual (block) time steps and the time step parameter $\eta$ was
set to $0.02$ \citep{1992PASJ...44..141M}. We also defined a maximum
time step of $2^{-5}$ and a minimum time step of $2^{-23}$ N-body time
units. A softening of $\epsilon = 0.01$ is used in both integration
methods.

We have performed two profiling experiments, using direct integration
whenever the separation of the central black holes was less than
$r_a$, and tree at all other times. The first experiment varies in the
number of simulation particles, while maintaining $r_a =
\sqrt{0.3}$. The other experiment uses 32k particles and a different
$r_a$ for each run. For comparison, we have also included a full tree
and a full direct run.

\subsection{Results}

\begin{figure}[t!]
  \centering
  \includegraphics[angle=270,scale=0.4]{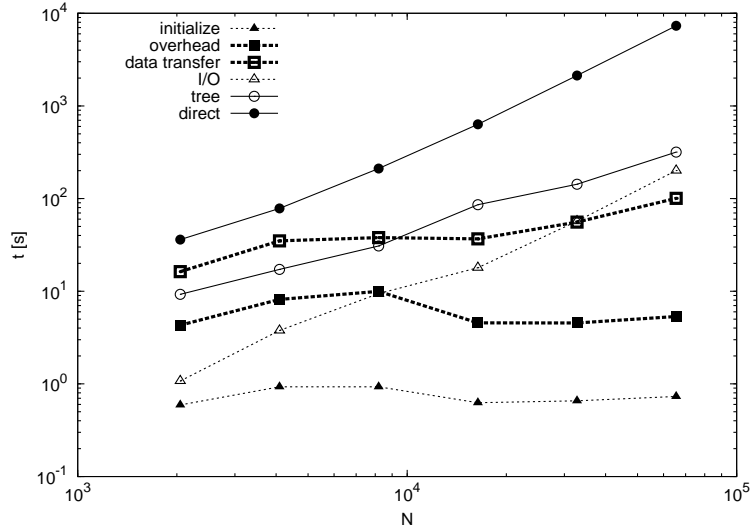}    

  \caption{Timing measurements of the living simulation tasks as a
    function of the number of simulated particles. The two solid lines
    represent time spent on direct integration (bullets) and tree
    integration (circles). The thick dashed lines indicate grid
    overhead by data transfers (open squares) and job submissions
    (filled squares). Finally, the two thin dashed lines indicate
    overhead caused by local file I/O (open triangles) and code
    initializations (filled triangles).}

\label{Fig:Overhead}
\end{figure}

\begin{figure}[t!]
  \centering
  \includegraphics[angle=270,scale=0.4]{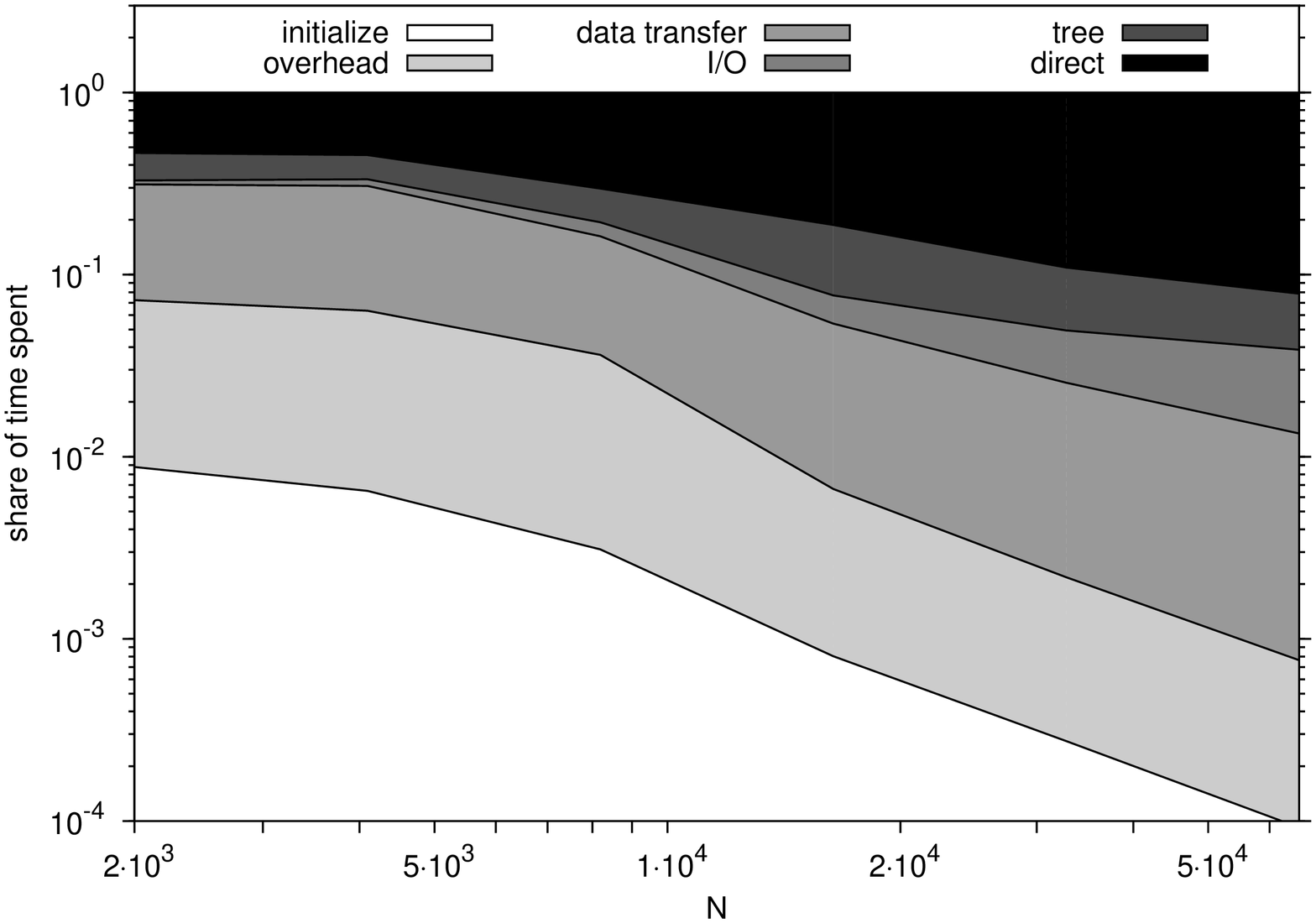}    

  \caption{Relative cumulative share of time spent by the living
    simulation tasks as a function of the number of simulated
    particles. From top to bottom the areas refer to the share of time
    spent on direct integration, tree integration, local file I/O,
    grid data transfer, grid job submissions and simulation
    initializations. Note that both axes are in log-scale.}

\label{Fig:RelOverhead}
\end{figure}

We have summarized the results of our living simulation
in two figures.  The absolute time spent on each task as a
function of the number of particles is given in
Fig.~\ref{Fig:Overhead}, and the relative time share of each task is
shown in Fig.~\ref{Fig:RelOverhead}. For all the tested initial
conditions, the simulation migrated itself three times,
resulting in four initializations and three simulation migrations per
run. 

In this experiment, we find that the direct N-body integration
dominates the simulation performance in all cases, and that for larger
$N$, the relative overhead caused by grid data transfers and job
submissions diminishes. Although the time spent on local I/O scales
steeply due to unoptimized identifier lookup calls (this has recently
been fixed in MUSE), this overhead remains relatively small throughout
our runs. When using 64k particles, we found that $\sim 4$ percent of
the simulation time is spent on overhead tasks.

We have performed several runs with 32k particles, using a different
$r_a$ for each run. The results of this experiment are shown in
Tab.~\ref{Table:switch16k}. During the runs we observed several close
interactions between the SMBHs, and a decreasing trend in the value of
$r_{\rm SMBH}$. This behavior caused the living simulation runs with
smaller $r_a$ to switch more frequently.

A pure tree integration ($r_a = 0$) leads to the highest cumulative
energy error, whereas a pure direct integration ($r_a = \infty$) has
the lowest error. When switching between both codes with the living
simulation, the energy error is lower than using pure tree, but much
higher than using a direct code. Even when using a $r_a = \sqrt{10}$,
where the code switches only once after 4 N-body time units, we see a
much larger error than when using only direct. The energy error is
dominated by the execution of the tree code. This difference is caused
by the tree-based force calculation as well as by the second-order
leapfrog integration scheme used in the tree code. A detailed
discussion on the energy behavior of these combined simulations can be
found in Harfst et. al. (2009, in preparation).

The simulation performance is dominated by N-body integration in all
cases, although there is a relatively high overhead for $r_a = 0.1$,
which is caused by the 29 switches. Each of these switches
  requires the particles to be saved locally, sent across the
Atlantic using regular internet, and loaded on the new machine. 

\section{Conclusion}
\label{sect:conc}

We introduced the living application as a way to manage complex
applications on a large distributed infrastructure.  Due to the
autonomous nature of a living simulation, it is important to provide a
mechanism that allows the user to terminate it. By having the
simulation retrieve its extended privileges from a credential
management service (MyProxy), users are able to revoke the privileges
of the simulation regardless of its location. In addition, we can
renew short-lived proxy credentials instead of using a long-lived
credential, which may be attractive to malicious users.

We then apply this concept in a living simulation of two galaxies
merging, using a straightforward and autonomous resource
  selection scheme which chooses from a predefined list of available
  resources. Our approach allows the simulation to use the optimal
compute resources for each of the two solvers, switching resources
whenever a different solver is required. In our example, the solvers
were a tree code and a direct $N$-body method, which were optimized
for two kinds of special-purpose hardware, namely a GPU (tree) and a
GRAPE (direct). The switches take place autonomously without user
intervention, remote output retrieval or external managers. In
  our experiments, the execution time was only affected marginally by
overhead such as caused by job migration and data transfer over the
grid. In the cases where each solver is best run on a
  different architecture and the overall simulation performance is not
  dominated by switching overhead, we find that the living simulation
  is a practical and resource efficient solution.

The creation of grid species enables us to give a simulation the
ability to autonomously use the grid, acquire and apply internal
knowledge, and migrate themselves. In this work we presented a
first implementation, which we intend to extend in the near
future. Possible extensions include connecting living applications
with grid resource monitoring and discovery services to dynamically
obtain information on resource availability, and developing a living
application which is able to recover from failures of grid nodes.
These extensions allow us to apply the living application to evolve to
a more complex organism, which can be applied to problems of 
greater complexity.

\section{Acknowledgements} 
We are grateful to Rick Quax for his contributions during the early
stages of this project, and Steve McMillan for technical support and
access to the GRAPE at Drexel University. We also thank David Groep
and Hashim Mohammed for their input on the security aspects of this work,
to Evghenii Gaburov for sharing his octgrav code for our experiments and
to Alfons Hoekstra for his very useful suggestions.

This research is supported by the Netherlands organization for
Scientific research (NWO) grant \#643.200.503, the European Commission
grant for the QosCosGrid project (grant number: FP6-2005-IST-5
033883), the Netherlands Advanced School for Astronomy (NOVA) and the
Leids Kerkhoven-Bosscha fonds (LKBF).

\section{Biographies}
Derek Groen has been a Ph.D. student at the University of Amsterdam in
the Section Computational Science as well as the Astronomical "Anton
Pannekoek" institute who recently moved to Leiden Observatory. He is
now a member of an interdisciplinary group that focuses on
high-performance simulation of astrophysical problems. Derek's
research interests are mainly in using high-performance, distributed
and dynamic grid computing to simulate astrophysical problems. Derek
obtained his M.S. in Grid Computing and his B.S. in Computer Science
at the University of Amsterdam.

Stefan Harfst is a post-doc at the University of Amsterdam, working in 
the Astronomical "Anton Pannekoek" Institute and in the Section 
Computational Science. He received his Ph.D. in Astrophysics from the 
University of Kiel, Germany, in 2005. After that, he was a post-doc at 
the Rochester Institute of Technology (Rochester, NY) for two years before 
moving to the Netherlands. Stefan's research interests are in computational 
astrophysics, mainly in stellar dynamics of dense stellar systems such as 
the Galactic center.  He has been collaborating on the development and 
performance analysis of high-performance N-body algorithms, which are 
now widely used within the N-body community. 

Simon Portegies Zwart received his Ph.D. in 1996 under the supervision
of prof. F. Verbunt at Utrecht University. He continued to work on
stellar clusters, first as JSPS fellow at Tokyo University and as
Hubble fellow at the Massachusetts Institute of Technology, after
which he returned to Amsterdam as a fellow of the Royal Netherlands
Academy of Sciences. He is currently a full professor on computational
astrophysics at Leiden Observatory in the Netherlands.  

\bibliographystyle{elsart-harv}

\bibliography{Library}

\pagebreak

\begin{table}
\centering

\caption{Specifications for the test nodes.  The first column gives
  the name of the computer followed by its country of residence (NL
  for the Netherlands, US for the United States). The subsequent
  columns give the type of processor in the node, followed by the
  amount of RAM, the operating system, and the special hardware
  installed on the PC.  Both nodes are connected to the internet
  with a 1\,Gbit/s Ethernet card.}

\begin{tabular}{lclrrrrr}
\hline
name     &location&CPU type    &RAM   &  OS    & hardware\\
         &        &            &[MB]  &        &\\
\hline
darkstar & NL &Core2Duo 3.0GHz & 2048 &Debian  & Nvidia 8800 Ultra \\
zonker   & US &2x Xeon 3.6GHz  & 2048 &Gentoo  & GRAPE 6A \\
\hline
\label{Tab:G3}
\end{tabular}
\end{table}

\pagebreak

\begin{table}[t!]
\centering
\caption{ Timing and energy measurements of the living simulation
  tasks using 32k particles with a different value $r_a$ during each
  run, given in the first column. The second column gives the number
  of switches during the simulation, while the subsequent columns
  respectively give the times spent on direct integration, tree
  integration and overhead tasks. The total execution time and the total relative energy error are
  respectively given in the last two columns.}
\begin{tabular}{lllllll}
\hline
$r_a$            &\# switches & direct    &tree   &other &total &$dE/E$ \\
                 &            & [s]       &[s]    &[s]   &[s]   & \\
\hline
0.0 (tree)       & 0          & 0         & 247   & 24   &271   &$1.47\cdot10^{-2}$\\
0.1              & 29         & 762       & 219   & 944  &1925  &$5.93\cdot10^{-3}$\\
$\sqrt{0.1}$     & 7          & 1820      & 160   & 257  &2237  &$3.54\cdot10^{-3}$\\
$\sqrt{0.3}$     & 3          & 2180      & 143   & 120  &2443  &$2.88\cdot10^{-3}$\\
1.0              & 3          & 2519      & 127   & 118  &2764  &$2.49\cdot10^{-3}$\\
$\sqrt{10}$      & 1          & 3624      & 64    & 54   &3742  &$1.04\cdot10^{-3}$\\
$\infty$ (direct)& 0          & 4528      & 0     & 5    &4533  &$2.77\cdot10^{-6}$\\
\hline
\label{Table:switch16k}
\end{tabular}
\end{table}

\end{document}